\documentclass[aps,pre,11pt,superscriptaddress,floatfix,tightenlines,showpacs,longbibliography,notitlepage]{revtex4-1}
\usepackage{graphicx}
\usepackage{bm}
\usepackage{amsfonts}
\usepackage{color}
\usepackage{amsmath}    

\usepackage{epsfig}
\usepackage{url}
\usepackage{hyperref}   
\usepackage{bm}
\usepackage{amssymb}
\AtBeginDocument{%
    \newwrite\bibnotes
    \def\bibnotesext{Notes.bib}
    \immediate\openout\bibnotes=\jobname\bibnotesext
    \immediate\write\bibnotes{@CONTROL{REVTEX41Control}}
    \immediate\write\bibnotes{@CONTROL{%
    apsrev41Control,author="08",editor="1",pages="1",title="0",year="1"}}
     \if@filesw
     \immediate\write\@auxout{\string\citation{apsrev41Control}}%
    \fi
}%

\newcommand{\req}{r_{\mathrm{eq}}}
\newcommand{\WiE}{\mathit{Wi}_E}
\newcommand{\WiB}{\mathit{Wi}_B}
\newcommand{\Qc}{\mathcal{Q}_c}

\newcommand{\ictsaddress}{International Centre for
  Theoretical Sciences, Tata Institute of Fundamental Research,
  Bangalore 560089, India}

\newcommand{\iitbaddress}{Department of Chemical Engineering, Indian Institute of Technology Bombay, Mumbai 400076, India}
  \newcommand{\niceaddress}{Universit\'{e} C\^{o}te d'Azur, CNRS, LJAD, Nice 06100, France}

\begin{document}
\title{Dynamics of a long chain in turbulent flows: Impact of vortices}
\author{Jason R. Picardo}
\email{jrpicardo@che.iitb.ac.in}
\affiliation{\iitbaddress}
\author{Rahul Singh}
\email{rahul.singh@icts.res.in}
\affiliation{\ictsaddress}
\author{Samriddhi Sankar Ray}
\email{samriddhisankarray@gmail.com}
\affiliation{\ictsaddress}
\author{Dario Vincenzi}
\email{dario.vincenzi@unice.fr}
\affiliation{\niceaddress}

\begin{abstract}

We show and explain how a long bead-spring chain, immersed in a homogeneous
isotropic turbulent flow, preferentially samples vortical flow
structures. We begin with an elastic, extensible chain which is
stretched out by the flow, up to inertial-range scales. This filamentary
object, which is known to preferentially sample the circular coherent
vortices of two-dimensional (2D) turbulence, is shown here to also
preferentially sample the intense, tubular, vortex filaments of 3D
turbulence. In the 2D case, the chain collapses into a tracer inside
vortices. In 3D, on the contrary, the chain is extended even in
vortical regions, which suggests that the chain follows axially-stretched
tubular vortices by aligning with their axes. This physical picture is
confirmed by examining the relative sampling behaviour of the
individual beads, and by additional studies on an inextensible chain
with adjustable bending-stiffness.  A highly-flexible, inextensible
chain also shows preferential sampling in 3D, provided it is longer
than the dissipation scale, but not much longer than the vortex tubes.
This is true also for 2D turbulence, where a long inextensible chain
can occupy vortices by coiling into them. When the chain is made
inflexible, however, coiling is prevented and the extent of
preferential sampling in 2D is considerably reduced.  In 3D, on the
contrary, bending stiffness has no effect, because the chain does not
need to coil in order to thread a vortex tube and align with its axis.

\end{abstract}


%

\maketitle

\section{Introduction}
\label{sect:introduction}

A three-dimensional (3D) incompressible
turbulent flow has a peculiar geometrical structure
which distinguishes it from a purely random field. 
The visualisation of the isosurfaces of enstrophy and energy dissipation
indeed indicates that vorticity concentrates into tubular structures,
while regions of intense strain are sheet-like~\cite{skh12} (see Fig.~\ref{fig:Qcube} for a visualization based on the ${\cal Q}$-criterion~\cite{Dubief,jlrs19}).
Several Lagrangian studies have~shown 
that the dynamics of objects
smaller than the viscous-dissipation scale can depend sensitively on the
nature of the local velocity gradient.
Heavy particles, for instance,
are ejected from vortical regions because of centrifugal forces
and thus concentrate in strain-dominated ones
\cite{maxey,se91}.
In turbulent channel flows,
highly stretched polymers are mainly found in the
regions of strong biaxial extension
that surround the near-wall streamwise vortices
\cite{sg03,terrapon04,bagheri12}.
Gyrotactic swimmers are trapped into the high-shear zones of a turbulent flow~\cite{boffetta16}, while ellipsoidal swimmers preferentially sample low-vorticity zones~\cite{Pujara2018}. 
Much less studied, however, is the case of an object whose size lies in the
inertial range of turbulence; its translation is coupled to its internal dynamics, which in turn is directly affected by coherent structures of the flow. 
This situation has only just begun to receive attention, especially in the context of flexible fibers\cite{verhille1,verhille2,gfv18,mazzino1,dc19,mazzino2}. 

A bead-spring chain is a simple physical system that allows us to investigate
the interaction between an extended filamentary object and the geometrical structure of a
turbulent flow. Such a system has been widely employed in polymer physics
\cite{de86}
and is here generalised in order to describe a chain longer than
the dissipation scale.
It consists of a sequence of beads connected
by phantom springs.
The extensibility of the chain can be tuned by varying
the strength and the maximum length of the springs, and its stiffness is
controlled by a bending potential that depends on the angle
between each neighbouring pair of springs. 
Even though a bead-spring chain can only be regarded as a very rudimentary model
of an elastic filament, it has the basic properties needed to
investigate the physical mechanisms that determine the motion of an
extensible and flexible object in a turbulent flow.
Moreover, its dynamics can be studied
in detail with moderate numerical effort.

In~\cite{picardo}, an extensible bead-spring chain was studied in a 
two-dimensional (2D), homogeneous, and isotropic, turbulent flow. It was shown 
that the centre of mass of the chain preferentially samples
the vortical regions. This is because the chain is strongly stretched
in high-strain regions and thus becomes unable to follow their evolution.
It eventually gets trapped into one of the large-scale vortices
that dominate a 2D turbulent flow. Inside the vortex, where straining is absent, it contracts 
and stays therein. Hence, in 2D turbulence, an extensible chain departs from straining regions but behaves like a tracer inside vortices, whence the
preferential sampling of the latter.

Here we study the dynamics of a bead-spring chain in 3D {turbulence, where}, unlike in 2D, vorticity-stretching leads to the formation of intense vortex tubes\cite{Moffat94,Davidson}, with significant straining along the tube axis.
We address the question of whether preferential sampling of vortical regions
persists in three dimensions and, if so, how it depends on the
extensibility and flexibility of the chain.

\section{Extensible chain}
\label{sect:extensible}

We consider a
chain consisting of $N_b$ identical inertialess beads {(see \cite{Singh-PRE} for a study 
of chains with inertial beads in two-dimensional turbulent flows)}, {each of which has a Stokes drag coefficient $\zeta$.
The beads are} connected to their nearest neighbours by nonlinear
springs, with equilibrium length $\req$,
maximum length $r_m$, and spring
coefficient $\kappa$.
The characteristic relaxation time of each spring is thus $\tau_E=\zeta/4 \kappa$, which in turn sets the relaxation time of the chain 
{$\tau_E^{\rm chain}=(N_b+1)N_b\tau_E/6$} \cite{collins}.
If $\bm x_i$, $1\leqslant i\leqslant N_b$, denote
the position vectors of the beads,
it is convenient
to describe the configuration of the chain in terms of the
position of its centre of mass, $\bm X_c=
(\sum_{i=1}^{N_b} \bm x_i)/N_b$, and the interbead separations 
$\bm r_j=\bm x_{j+1}-\bm x_j$, $1\leqslant j\leqslant N_b-1$.

We first consider a freely-jointed
(the links do not oppose resistance to bending),
extensible chain, as in~\cite{picardo}.
The equations of motion for such a chain are:
\begin{subequations}
\begin{align}
  \dot{\bm X}_c&=\dfrac{1}{N_b}\sum_{i=1}^{N_{\rm b}} \bm u(\bm x_i,t)
  +
  \dfrac{1}{N_b}\sqrt{\dfrac{\req^2}{6\tau_E}}\sum_{i=1}^{N_{\rm b}} \bm\xi_i(t),
  \label{eq:cm-ext}\\
  \dot{\bm r}_j&=\bm u(\bm x_{j+1},t)-\bm u(\bm x_j,t)
  +\dfrac{1}{\zeta}\left(\bm f^E_{j+1}-\bm f^E_j\right)
  +\sqrt{\dfrac{\req^2}{6\tau_E}}\,[\bm\xi_{j+1}(t)-\bm\xi_j(t)],
\qquad (1\leqslant j\leqslant N_b-1).
\label{eq:disp-ext}
\end{align}%
\label{eq:motion-ext}%
\end{subequations}%
The elastic force on the $j$-th bead, $\bm f^E_j$, takes the form:
\begin{equation}
\bm f^E_j=\kappa\,\alpha_j\hat{\bm r}_j
\dfrac{r_j}{{1-r_j^2/r_m^2}}
-\kappa\,\alpha_{j-1}\hat{\bm r}_{j-1}
\dfrac{r_{j-1}}{{1-r_{j-1}^2/r_m^2}},
\end{equation}
with $r_j=\vert\bm r_j\vert$, $\hat{\bm r}_j=\bm r_j/r_j$ and
\begin{equation}
  \alpha_j=
  \begin{cases}
    0 & \text{if $j\leqslant 0$ or $j=N_b$}
    \\
    1 & \text{otherwise.}
  \end{cases}
  \label{eq:alpha}
\end{equation}
The divergence of the force at $r=r_m$ ensures that the
length of the chain, $R=\sum_{j=1}^{N_b-1}r_j$, stays
smaller than the maximum value $L_m=(N_b-1)r_m$.
The vectors $\bm\xi_j(t)$, $1\leqslant j\leqslant N_b$,
are independent multi-dimensional
white noises; they serve the purpose of modeling the
collisions between the beads and the molecules of the fluid, thereby
setting the equilibrium end-to-end length of the chain to
$\req\sqrt{N_b-1}$.
Equations~\eqref{eq:motion-ext} generalise the well-known 
Rouse model of polymer physics \cite{de86} in such a way as to 
account for the nonlinearity of the velocity field.
Indeed, in the original Rouse model the velocity differences
between the beads are replaced by their first-order Taylor expansion
in the separation vectors, because polymers are assumed to be
much shorter than the smallest scale of the velocity field.
Here the full 
velocity differences are retained, since the chain is allowed to extend into the 
inertial range of turbulence.
An analogous generalisation was applied 
in~\cite{mazzino-3,pg03,ds06,av16}
to the elastic dumbbell model ($N_b=2$).

\begin{figure}[t]
\centering\includegraphics[width=0.7\textwidth]{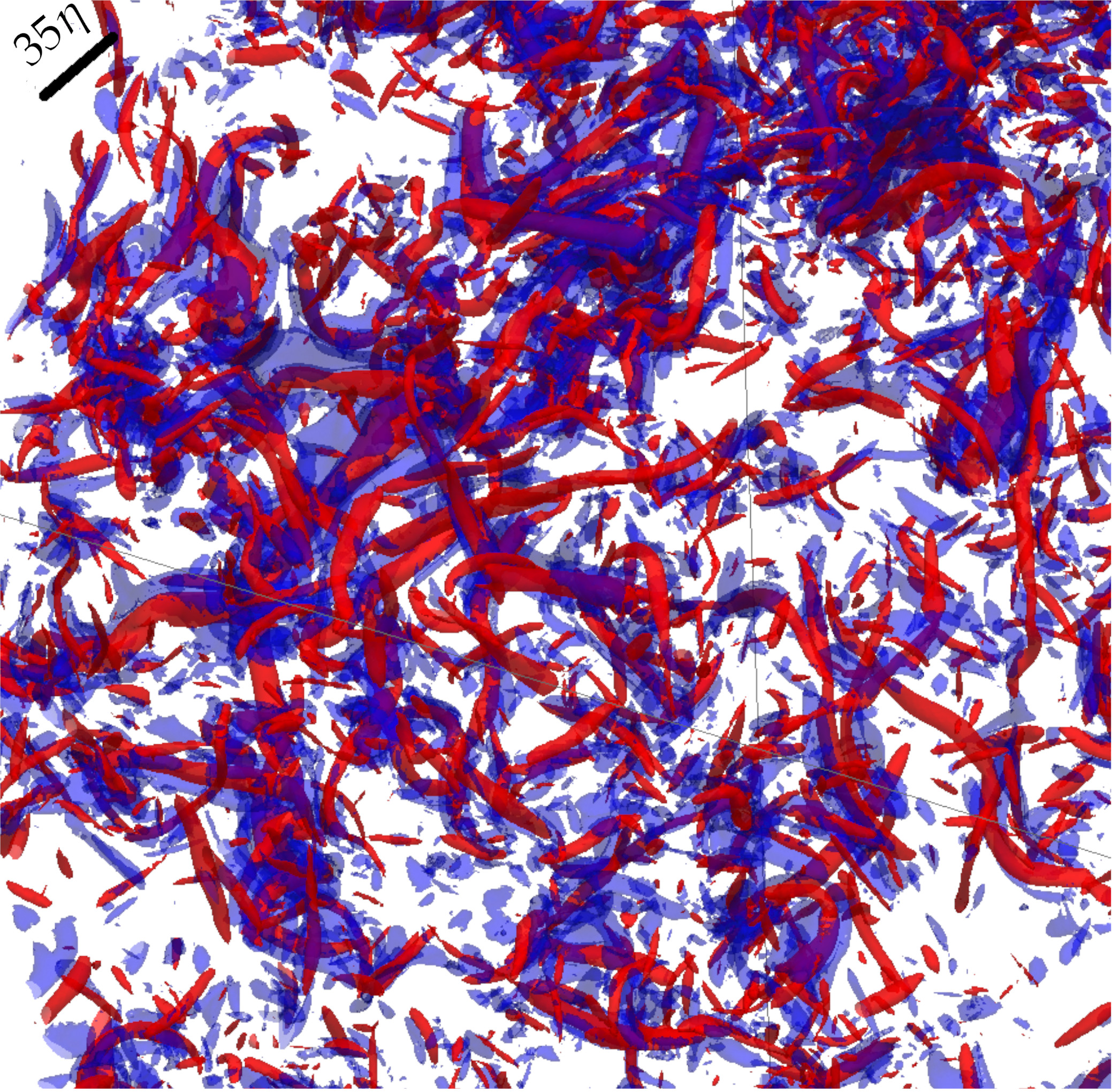}
\caption{Contours of ${\cal Q}$ showing intense, rotational, vortex tubes (red, ${\cal Q} = +5 \sqrt{ \langle {\cal Q}^2 \rangle}$) enveloped by strong straining sheets (blue, ${\cal Q} = -2 \sqrt{ \langle {\cal Q}^2 \rangle}$), in three-dimensional homogeneous isotropic turbulence. The black scale bar at the top-left corner corresponds to a length of $35\eta$.}
\label{fig:Qcube}
\end{figure}

Finally, the velocity field $\bm u(\bm x,t)$ describes
the motion of the fluid and is the solution of the incompressible
Navier--Stokes equations,
\begin{equation}
\partial_t\bm u+\bm u\cdot\nabla\bm u=-\nabla p+\nu\Delta\bm u+\bm F,
\qquad \nabla\cdot\bm u=0,
\label{eq:NS}
\end{equation}
where $p$ is pressure,
$\nu$ the kinematic viscosity of the fluid, and $\bm F(\bm x,t)$ is a large-scale body forcing which maintains a homogeneous, {isotropic, and} statistically stationary turbulent flow. We perform direct numerical simulations (DNS) on a $2\pi$-periodic domain, using a standard de-aliased pseudo-spectral method. For 3D simulations, we use a $N^3=512^3$ spatial grid and a second order Adams-Bashforth time-integration scheme. The body-forcing $\bm F(\bm x,t)$ injects a constant amount of energy---equalling the mean dissipation rate $\epsilon$---into the first two wave-number shells. The Kolmogorov dissipation time and length scales are given by $\tau_{\eta} = (\nu/\epsilon)^{1/2}$ and  $\eta = (\nu^3/\epsilon)^{1/4} \approx 1.7 \, k_{\rm max}$ (where $k_{\rm max}=\sqrt{2}N/3$ is the maximum resolved wavenumber). The results presented below correspond to a flow with Taylor-Reynolds number $Re_{\lambda} = 2 E \sqrt{5/(3 \nu \epsilon)}=196$ (where $E$ is the mean kinetic energy), which is sufficiently large for a clear inertial range (albeit less than a decade) to emerge in the energy {spectrum and} for the formation of distinct vortex tubes (cf. Fig.~\ref{fig:Qcube}). We have checked that the sampling behaviour we describe persists even for smaller $Re_{\lambda}$ of $123$ and $64$ as well, though it does intensify with $Re_{\lambda}$ over this limited range.

While our primary focus is on chains in a 3D turbulent flow, it is helpful to compare their 3D dynamics with that in a 2D turbulent flow, especially when addressing the effects of extensibility and flexibility. For 2D flow simulations, a $N^2=1024^2$ spatial grid and a second order Runge-Kutta time-integration scheme is used, as in Ref. ~\cite{picardo}. A constant-in-time forcing ${\bm F} = F_0 \,{\rm sin} (k_f x)\, {\bm e_y}$ is applied, where $F_0$ is the forcing amplitude and $k_f = 5$ is the forcing wavenumber, which sets the scale of the large coherent vortices, $2 \pi k_f^{-1}$. An Ekman friction term~\cite{Prasad2D,Boffetta2D} with coefficient $\mu = 10^{-2}$ is included in (\ref{eq:NS}) to damp out the energy at the large scales (due to an inverse cascade).

After the flow (3D or 2D) attains a stationary state, we introduce a large number of chains, of order $10^4$, each with $ N_b=20 $ beads. We evolve many chains simultaneously only to obtain good statistics; the dynamics we study are of a single chain in the flow. The chains are given a small initial length, close to the no-flow equilibrium value. They are then allowed to be stretched out by the flow and attain a steady-state distribution before we begin recording statistics. The equations governing the dynamics of the {centre of mass} and separation vectors of the chain are integrated using a second-order Runge-Kutta scheme, augmented by a rejection algorithm~\cite{Ottinger} that prevents the nonlinear spring force from diverging as $|\bm r|$ approaches $r_m$. The chains are evolved with a time step {$\Delta t_{\rm chain} = \Delta t_f/N_{\rm sub}$}, where $\Delta t_f$ is the time-step of the fluid flow solver and $N_{sub}$ is the number of sub-steps taken by the chain solver for every step of the fluid solver. The flow is assumed to be unchanging over the duration of the sub-steps. While $N_{\rm sub}$ is set to unity for the freely-jointed chain, used in this section, we use up to $N_{\rm sub}=10^3$ for accurately resolving the numerically-stiff dynamics of inextensible and inflexible chains, to be introduced later in \S~\ref{sec:inextensible} and \ref{sec:flexible}.

The influence of elasticity on the dynamics of the chain is described in terms of the elastic Weissenberg number {$\WiE=\tau_E^{\rm chain}/\tau_f$}, which is the ratio of the chain relaxation time to the viscous-dissipation {time scale} of the flow. For 3D turbulence, $\tau_f=\tau_\eta$, whereas for 2D flows we choose the small {time scale} associated with the dissipation of enstrophy, $\tau_f = \langle 2 \omega^2 \rangle^{-1/2}$, where $\langle \omega^2 \rangle$ is the mean enstrophy. For small $\WiE$, the chain is in a contracted configuration and acts like a tracer; for large $\WiE$, it is stretched out by the flow.

As mentioned in \S~\ref{sect:introduction}, the dynamics of an extensible chain
was studied in \cite{picardo} for a
2D turbulent flow forced at large spatial scales. The equilibrium size
of the chain was assumed to be of the order of the dissipation scale, while
$L_m$ was much longer and comparable to the scale of the coherent vortices, $2 \pi k_f^{-1}$,
or even greater than it.
When $\WiE$ was sufficiently large for the typical length of the chain to
approach the size of the vortices, then the chain was shown to exhibit a marked preferential sampling of
vortical regions. The mechanism that was proposed in \cite{picardo}
to explain this phenomenon, already
sketched briefly in \S~\ref{sect:introduction}, can be summarised as follows:
\begin{enumerate}
\item Inside a vortex, where stretching is absent, the chain shrinks down to its equilibrium size.
  Since the equilibrium size of the chain is much smaller than the
  size of the vortex, the chain essentially behaves as a tracer and follows
  the vortex during its lifetime. 
\item In a region of intense strain, the chain is stretched to the extent 
that it cannot continue to follow the
  straining region. The chain eventually encounters a vortex that coils it up
  and `entraps' it according to the dynamics described above.
\end{enumerate}
The combination of these effects causes an extensible chain to stay
inside large vortices and leave high-strain regions, which results in
a strong preferential sampling of vortices.
This phenomenon was quantified in \cite{picardo} by studying
the Okubo-Weiss parameter \cite{okubo,weiss} evaluated at the position of the centre
of mass of the chain,
\begin{equation}
  \Lambda_c=\dfrac{\omega_c^2-\sigma_c^2}{4\langle\omega^2\rangle},
  \label{eq:OW}
\end{equation}
where $\omega_c$ and $\sigma_c$ 
are the vorticity and the strain rate
at the position of the centre of mass, respectively.
We recall that positive values of $\Lambda_c$ correspond to vorticity-dominated
regions, whereas negative values indicate strain-dominated regions.
For large enough values of $\WiE$,
the probability of positive values of $\Lambda_c$
was found to be
much higher than for a tracer transported by the same flow \cite{picardo}.
In addition, the joint probability density function (PDF) of {the chain length} $R$ and $\Lambda_c$
confirmed that  the chain
is contracted in vortical regions and extended in straining ones.

\begin{figure}[t]
\centering\includegraphics[width=\textwidth]{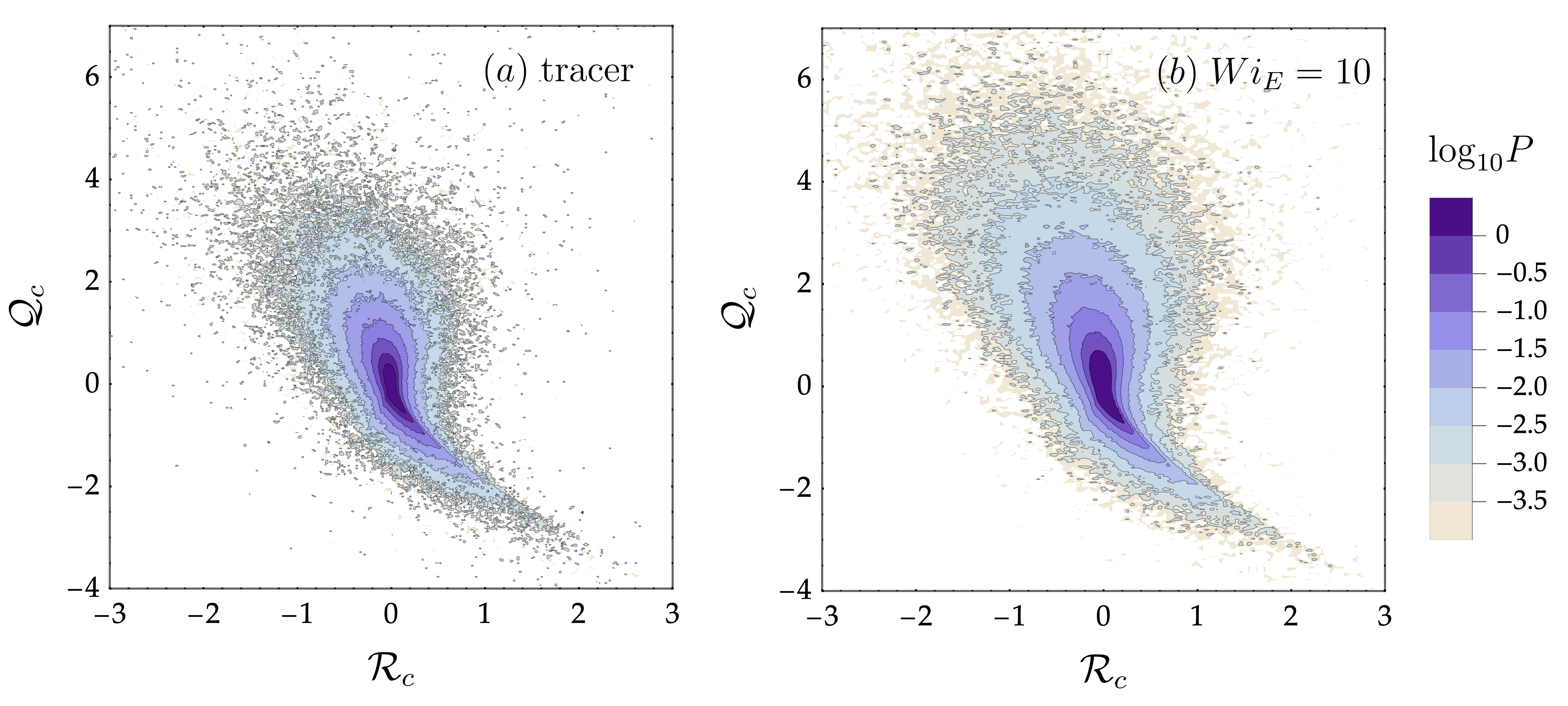}
\caption{Joint PDF of ${\cal R}_c$ and ${\cal Q}_c$ (\textit{a}) for a tracer 
and (\textit{b}) for an inertialess, freely-jointed,
extensible chain with $\WiE=10$, $L_m=40\eta$, $\req=0.045 \eta$.
}
\label{fig:QR}
\end{figure}

The first question we address here is whether or not an extensible chain
exhibits preferential sampling of vortical regions also in 
a 3D turbulent flow. A substantial difference indeed exists between
2D and 3D flows. A 2D turbulent flow is characterised by
large, long-lived vortices, inside which stretching is weak.
In a 3D turbulent flow, vorticity concentrates into intense tubular structures with 
significant stretching along their axes. The dynamics of the chain
in vorticity-dominated regions is therefore expected to be substantially
different in the two cases.

In three dimensions,
the local nature of the flow can be classified by
using the $\cal Q$-$\cal R$ representation of the
velocity gradient \cite{cantwell96}. 
If ${\sf A}_c=\tau_\eta\nabla\bm u(\bm X_c(t),t)$ denotes
the rescaled velocity gradient at the position of the centre of mass,
let ${\cal Q}_c=-\operatorname{tr}\mathsf{A}_c^2/2$ 
and ${\cal R}_c=-\operatorname{det}\mathsf{A}_c$ be its second and third
invariants.
Since ${\cal Q}_c$ can be rewritten as 
\begin{equation}
\mathcal{Q}_c=\tau_\eta^2\frac{\omega_c^2/2-\mathsf{S}_c:\mathsf{S}_c}{2}; \;\; 
\mathsf{S}_c = (\mathsf{A}_c+\mathsf{A}_c^{\rm T})/2,
\end{equation}
the sign of ${\cal Q}_c$ discriminates between the regions dominated by
vorticity ($\mathcal{Q}_c>0$) and those dominated by strain
($\mathcal{Q}_c<0$). Thus, $\mathcal{Q}_c$ has a role analogous to that played by $\Lambda_c$ in a 2D flow. Figure~\ref{fig:Qcube} presents a snapshot of the iso-surfaces of $\cal Q$ from our 3D simulation, which clearly reveal sheet-like straining zones (large negative $\cal Q$ in blue) in close proximity to intense tubular vortices (large postive $\cal Q$ in red).

Figure~\ref{fig:QR} compares the steady-state joint PDFs of $\mathcal{R}_c$
and $\mathcal{Q}_c$ for a tracer particle (panel {\it a}) and for the extensible chain with a large $\WiE = 10$ (panel {\it b}). The chain has an equilibrium size of about $0.4 \eta$ and a maximum extension $L_m=40\eta$, so that when the chain is stretched {its length} $R$ lies in
the inertial range and is comparable to the typical size of vortex tubes in our simulations (cf. the vortex tubes and scale bar in Figure~\ref{fig:Qcube}). The shape of the joint PDF of the chain is similar to 
that of the tracer. However, the probability of sampling positive  $\Qc$ is significantly
larger for the chain, than for the tracer.
Therefore, we deduce that
preferential sampling of vorticity-dominated regions persists in three
dimensions.

As mentioned above, however, a major difference exists between the chain
dynamics in 2D and 3D turbulence.
\begin{figure}[t]
\centering\includegraphics[width=0.95\textwidth]{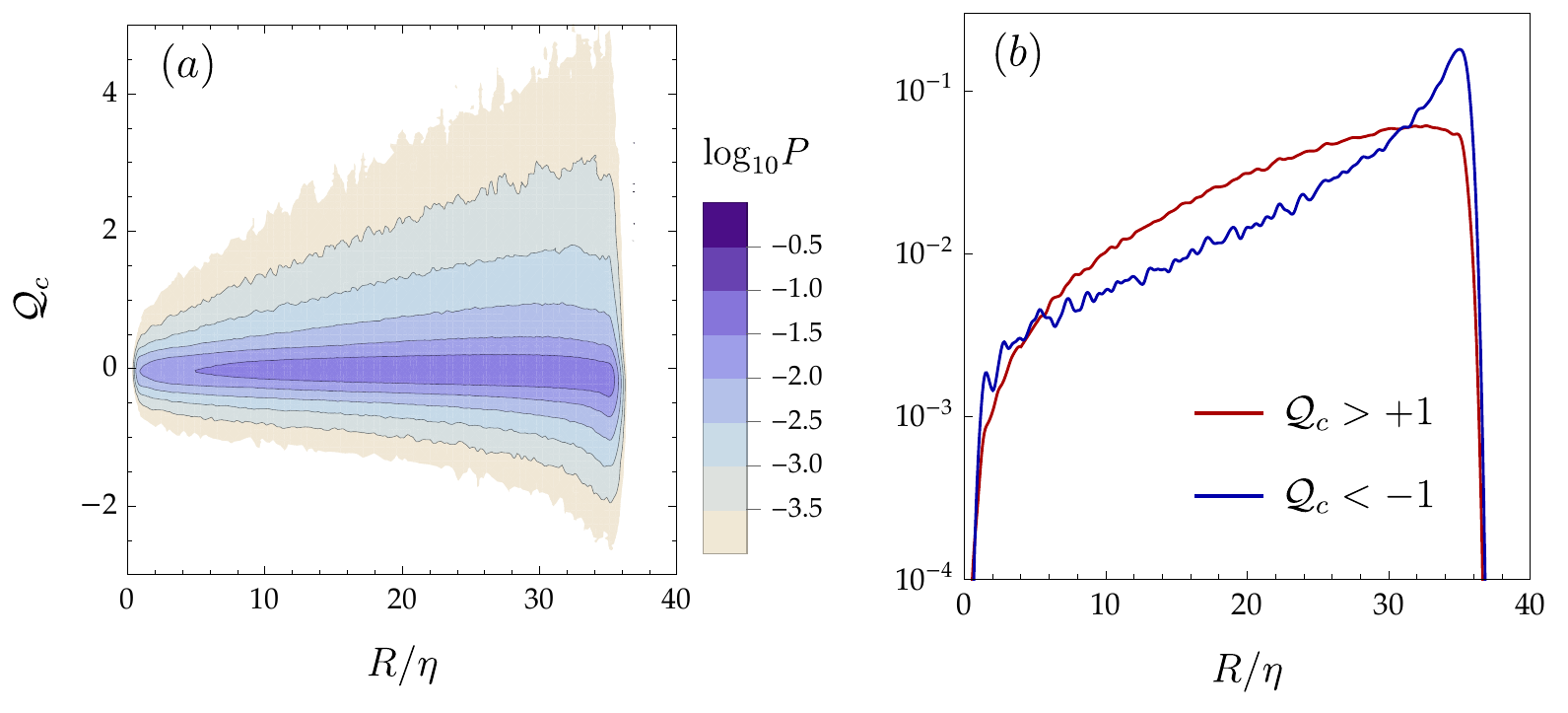}
\caption{(\textit{a}) Joint PDF of the chain length
  $R$ (rescaled by the dissipation scale $\eta$) and $\Qc$ for a
freely-jointed inextensible chain.
  (\textit{b}) Conditional PDF of $R$ rescaled by $\eta$ given $\Qc<-1$
  (blue curve) and $\Qc>1$ (red curve).
The parameters are $\WiE=10$, $L_m=40\eta$, $\req=0.045 \eta$ 
in both panels.}
\label{fig:extension}
\end{figure}
This is reflected in 
the conditional PDF of the chain length $R$ given the value of $\Qc$, presented in figure~\ref{fig:extension}{\it b}, which
shows that contrary to the 2D case
the chain is considerably stretched not only
in strain-dominated but also in vorticity-dominated regions.
This difference is clear from the 
comparison of the joint PDF of $R$ and $\Qc$, shown in figure~\ref{fig:extension}\textit{a}, with
its 2D analog (see figure~3 in \cite{picardo}) and has important 
consequences for preferential sampling.
In two dimensions, indeed,
the contraction of the chain inside vortices was 
identified as an essential element of the sampling dynamics \cite{picardo} (though we shall revisit this idea in section~\ref{sec:flexible}).
In three dimensions, the chain
does not collapse into a tracer
even when the flow has a strong vortical nature (figure~\ref{fig:extension}\textit{b}). Hence,
the mechanism leading to preferential sampling
ought to be different from that operating in 2D turbulence.

Now, the only way an elongated chain can remain inside a vortex tube of comparable length,
is if it aligns itself along the vortex axis. Given that
the majority of vortices are axially stretched (figure~\ref{fig:QR}), this scenario is consistent with our observation that
the chain is typically stretched out inside a vortex  (figure~\ref{fig:extension}\textit{b}). In contrast, it is unlikely that the entire chain can be encapsulated into straining regions, given their less coherent, sheet-like topology. Moreover, the evolution of straining regions is much more non-local in nature than that of the vortex tubes~\cite{Tsinober}, which tend to move with the fluid, except for the effects of viscous diffusion (in the inviscid limit the tubes would be 'frozen' into the fluid as a consequence of Kelvin's circulation theorem~\cite{Davidson}). Thus, we expect a chain to be able to enter and follow a vortex tube more easily than a straining region. 

\begin{figure}[t]
\centering\includegraphics[width=0.9\textwidth]{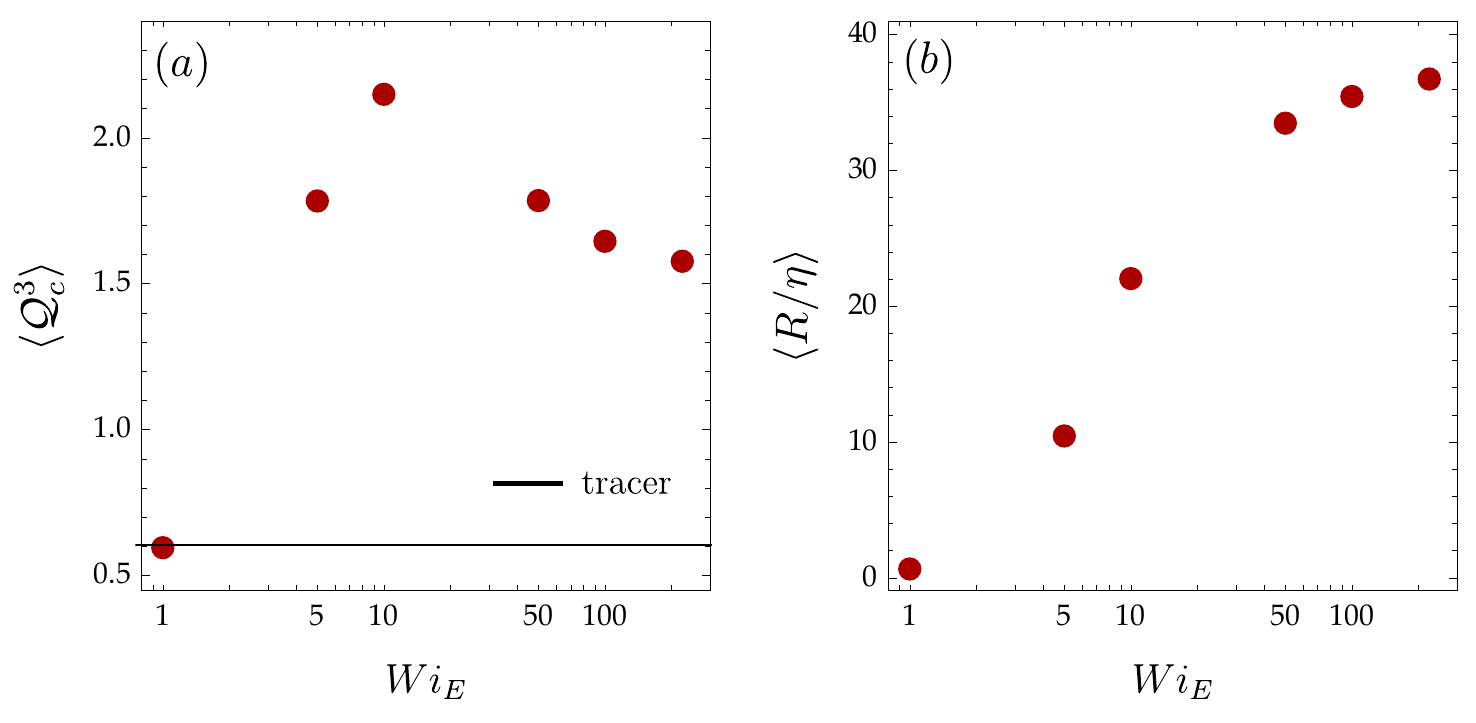}
	\caption{(\textit{a}) Third moment of the distribution of ${\cal Q}$ and (\textit{b}) the average chain length as a function of $\WiE$, for fixed 
	$L_m=40\eta$ and $\req=0.045 \eta$. The value of $\langle \Qc^3 \rangle$ for a tracer is shown as a horizontal black line in panel {\it a}; it is {positive} due to the presence of intense vortex tubes. 
}
\label{fig:Q_R_Wi}
\end{figure}

This explanation implies that there is an ideal chain elasticity for preferential sampling: If $\WiE$ is too small then the chain acts like a tracer, whereas, if $\WiE$ is too large then the chain may stretch out beyond the length of the vortex tubes and be unable to reside inside them. This intuition is corroborated by figure~\ref{fig:Q_R_Wi}, which shows the effect of $\WiE$ on preferential sampling (panel {\it a}), as well as on the mean length of the chain (panel {\it b}). We find that the third moment of the PDF of ${\cal Q}_c$, whose positivity implies a higher probability of positive $\Qc$, varies non-monotonically with $\WiE$, peaking at a value which corresponds to a mean chain length $\langle R \rangle \approx 20 \eta$. For larger $\WiE$, all the chains approach the maximum chain length $L_m = 40 \eta$ and cannot reside within the vortex tubes as effectively (cf. figure~\ref{fig:Qcube}).

We shall present more direct evidence for our explanation of 3D preferential sampling in section~\ref{sec:flexible}. But first, let us address the question that naturally arises from figure~\ref{fig:extension}\textit{a}: Is elasticity essential for preferential sampling, given that the chains are stretched out even inside vortices? The physical picture of chains aligning along vortex tubes would hold even for long chains of a fixed length, and so we would expect such inextensible chains to show preferential sampling as well. We test this idea in the next section.
%

\section{Inextensible chain}
\label{sec:inextensible}

In order to describe a `macroscopic' inextensible chain, we disregard
Brownian fluctuations in \eqref{eq:motion-ext}
and replace~$\bm f^E_j$ with
\begin{equation}
\tilde{\bm f}^E_j=\kappa\,\alpha_j\hat{\bm r}_j
(r_j-\req)
-\kappa\,\alpha_{j-1}\hat{\bm r}_{j-1}
(r_{j-1}-\req),
\end{equation}
where $\req$ still has the meaning of 
equilibrium length of the springs.
$\WiE$ is set to a very small value
to ensure that the chain is in effect inextensible.
Its length is then $L=(N_b-1)\req$
(in the simulations, we set $\WiE = 0.1$, which ensures that {$R=\sum_{j=1}^{N_b-1}r_j$} differs from $L$
by $2\%$ at most).

\begin{figure}[!t]
\centering\includegraphics[width=0.9\textwidth]{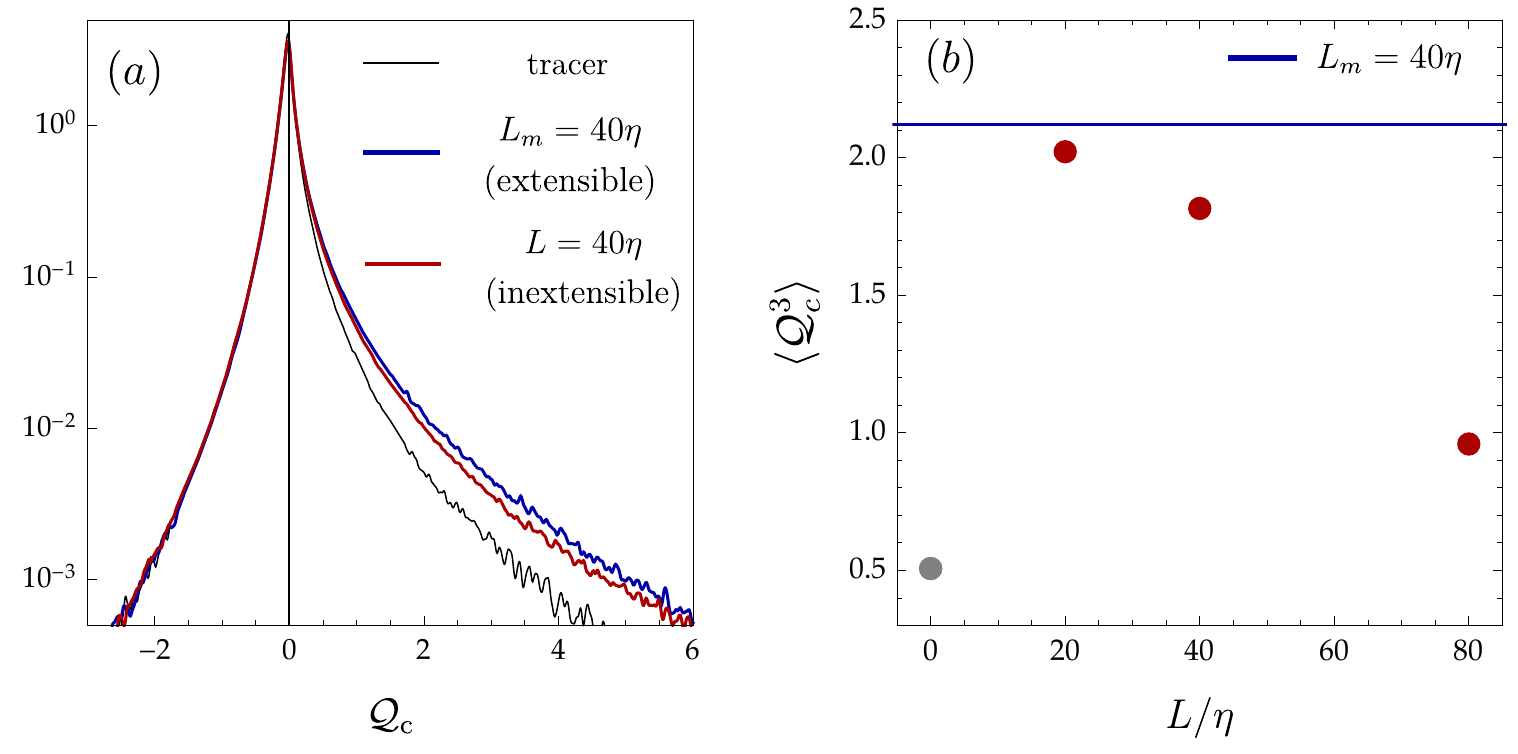}
\caption{PDF of $\Qc$ for a tracer (black curve),
an extensible chain with $\WiE = 10$, $\req=0.045\eta$ and $L_m=40\eta$ (blue curve), and
an inextensible chain with $L=L_m$ (red curve).
(\textit{b}) Third moment of the PDF of $\Qc$
for a tracer (grey bullet) and for an inextensible chain
with various lengths $L$ (red bullets). The blue line
indicates the value of $\langle\Qc^3\rangle$ for the extensible {chain} of panel {\it a}. The mean length of
this extensible chain is close to $20 \eta$, as seen in figure~\ref{fig:Q_R_Wi}{\it b} (case of $\WiE=10$).
}
\label{fig:inextensible}
\end{figure}

We now consider an inextensible chain having the same
length as the maximum length of the extensible chain (with $\WiE = 10$)
considered in \S~\ref{sect:extensible}, \textit{i.e.}
we take $L=L_m=40 \eta$.
The PDFs of $\Qc$ for the two cases
are compared in figure~\ref{fig:inextensible}\textit{a}.
The absence of extensibility is seen to have only a small effect.

The influence of the length $L$ of the inextensible chain on the level of preferential sampling
 is depicted in figure~\ref{fig:inextensible}\textit{b}, in terms of the third moment of $\Qc$. Here, we see the same non-monotonic variation of ${\langle \Qc^3 \rangle}$ with $L$, that we saw with $\WiE$ for the extensible chain (figure~\ref{fig:Q_R_Wi}\textit{a}). Moreover, the value of ${\langle \Qc^3 \rangle}$ for $L=20 \eta$ is very close to that of the extensible chain (horizontal line in panel \textit{b}), which has a mean length $\langle R \rangle \approx 20 \eta$ (case of $\WiE = 10$ in figure~\ref{fig:Q_R_Wi}\textit{b}). These results clearly demonstrate that preferential sampling
 persists even when the chain is not collapsible. Indeed the only role of extensibility is to allow a chain with a short equilibrium length to be stretched by the flow and
reach the appropriate length for staying in a vortex; it 
does not influence the preferential-sampling dynamics otherwise. This finding supports our
intuition regarding the mechanism of preferential sampling whereby
the chain resides in intense vortical regions by aligning with
the axis of vortex tubes.
Such a mechanism, indeed, does not rely on the
extensibility of the chain, and is equally applicable to inextensible, but long, chains.

The results in figure~\ref{fig:inextensible}{\it b} also confirm our understanding of figure~\ref{fig:Q_R_Wi}{\it b}, that there is an optimal 
length for preferential sampling---either attained dynamically by stretching or permanently fixed from the outset---that is comparable
to the characteristic linear size of vortex tubes.

\section{Inflexible chain}
\label{sec:flexible}
So far we have considered a chain that does not oppose bending. However,
taking the bending stiffness of the chain into account allows us to
further understand preferential sampling in 3D turbulence. Specifically, by increasing
the stiffness of the chain, we can check whether it is necessary for chains to be able to coil in order to enter vortex tubes. Coiling certainly plays an important role in preferential sampling in 2D turbulence, as demonstrated indirectly in \cite{picardo}, where the deformability of the chain was controlled by changing {the number of beads} $N_b$ while keeping {the maximum length} $L_m$ fixed. Here, we directly incorporate the forces arising from bending stiffness into the equations of motion, by assuming that the chain has a bending energy given by \cite{gs06}
\begin{equation}
E^B=A\req^{-1}\sum_{j=2}^{N_b-1}
(1-\hat{\bm r}_j\cdot\hat{\bm r}_{j-1}),
\end{equation}
where $A>0$ determines the bending stiffness. (Note that our definition of the separation vectors $\bm r_j$
differs from that used in~\cite{gs06}.)
The bending energy generates a force that depends on the angle between
two neighbouring links and whose effect is to restore the chain into a
rod-like configuration.
The form of the force on the $j$-th bead is \cite{gs06}
\begin{multline}
\label{eq:bending}
\bm f_j^B=\dfrac{A}{r_{eq}}\left[
\frac{\alpha_{j-2}}{r_{j-1}}\,\hat{\bm r}_{j-2}-
\left(\dfrac{\alpha_{j-2}}{r_{j-1}}\,\hat{\bm r}_{j-2}\cdot \hat{\bm r}_{j-1}
+\frac{\alpha_{j-1}}{r_{j}}
+\dfrac{\alpha_{j-1}}{r_{j-1}}\,\hat{\bm r}_{j-1}\cdot\hat{\bm r}_{j}
\right)\hat{\bm r}_{j-1}
\right.
\\
\left.
+\left(\dfrac{\alpha_{j-1}}{r_{j}}\,\hat{\bm r}_{j-1}\cdot \hat{\bm r}_{j}
+\frac{\alpha_{j-1}}{r_{j-1}}
+\dfrac{\alpha_{j}}{r_{j}}\,\hat{\bm r}_{j}\cdot\hat{\bm r}_{j+1}
\right)\hat{\bm r}_{j}
-\frac{\alpha_{j}}{r_{j}}\hat{\bm r}_{j+1}
\right],
\end{multline}
where $\alpha_j$ was defined in \eqref{eq:alpha}.
The characteristic time associated with this force is
$\tau_B=\zeta \req^3/A$,
and a dimensionless
measure of it is the `bending' Weissenberg number
$\WiB=\tau_B/\tau_\eta$.
The chain is inflexible and rod-like for small $\WiB$, while the
freely-jointed limit is recovered for large $\WiB$.

\begin{figure}[!t]
\centering\includegraphics[width=0.9\textwidth]{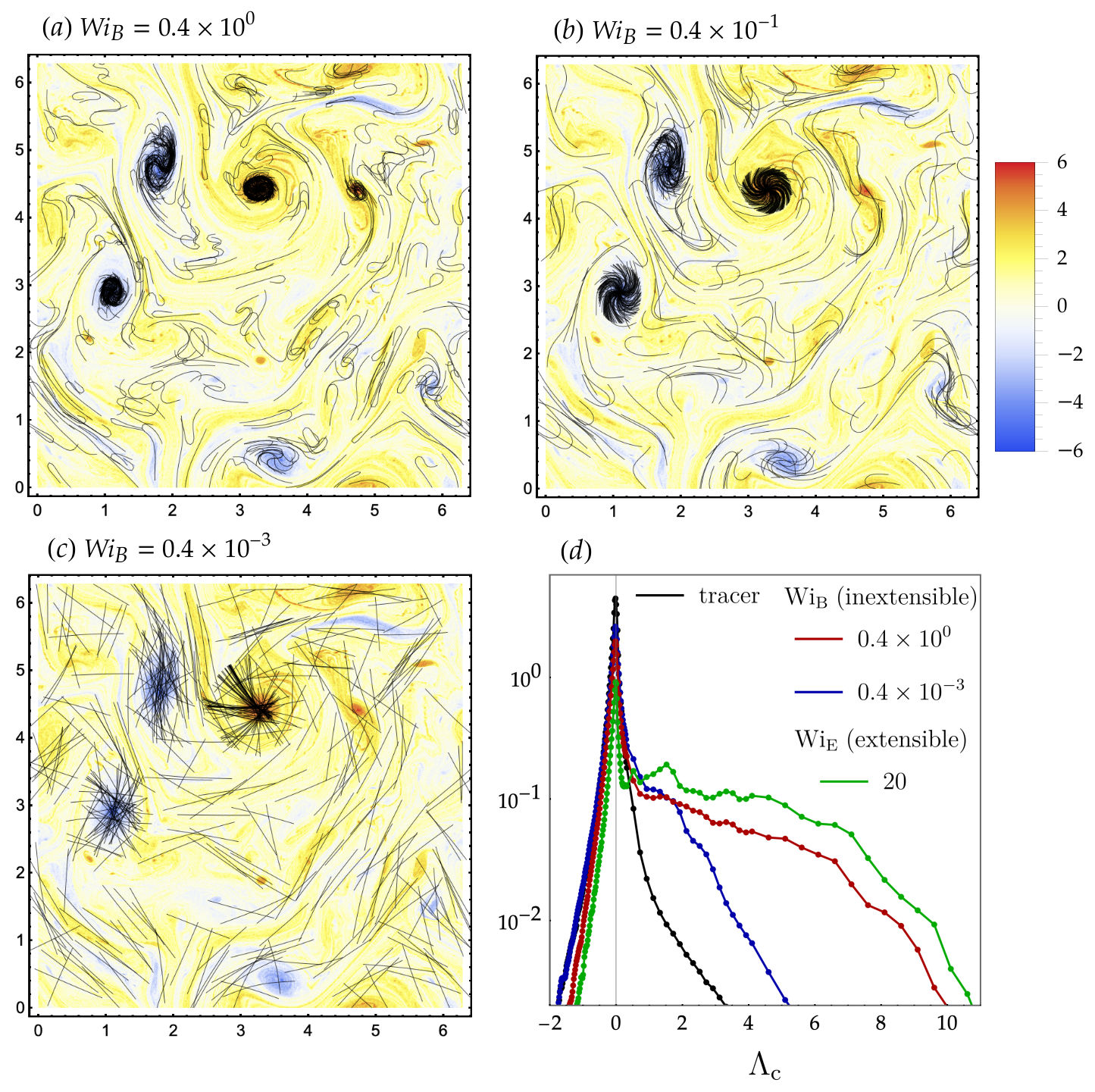}
\caption{Snapshots of an ensemble of inextensible
chains (black lines) in a 2D turbulent flow, overlaid on the vorticity field, corresponding to ({\it a}) $\WiB = 0.4$ (highly flexible), ({\it b}) $\WiB = 0.4 \times 10^{-2}$ (moderately flexible) and ({\it c}) $\WiB = 0.4 \times 10^{-3}$ (inflexible).
It should be noted that several chains are plotted at the same
time in order for the reader to visualise the preferential sampling
of vortices more easily. There is, however, no interaction between
the chains. The length of the chain $L$ is equal to the large vortex scale $2\pi/k_f =1.25$.
(c) PDF of the Okubo--Weiss parameter for the inextensible chain of panels {\it a} and {\it c}, along with the PDF for a tracer, as well as that for an extensible chain ($\WiE = 20$) with equilibrium size of $0.04$ and maximum length $L_m$ equal to the length $L$ of the inextensible chain.}
\label{fig:2D}
\end{figure}
%

With the addition of the bending stiffness,
the evolution equations for an inertialess, inextensible
chain become:
\begin{subequations}
\begin{align}
\dot{\bm X}_c&=\dfrac{1}{N_b}\sum_{i=1}^{N_{\rm b}} \bm u(\bm x_i,t),
\label{eq:cm}\\
\dot{\bm r}_j&=\bm u(\bm x_{j+1},t)-\bm u(\bm x_j,t)
+\dfrac{1}{\zeta}\left(\tilde{\bm f}^E_{j+1}-\tilde{\bm f}^E_j\right)
+\dfrac{1}{\zeta}\left(\bm f^B_{j+1}-\bm f^B_j\right)
\qquad (1\leqslant j\leqslant N_b-1).
\label{eq:disp}
\end{align}%
\label{eq:motion}%
\end{subequations}%

We begin by assessing the effect of bending stiffness on preferential sampling in two dimensions. As we already appreciate the importance of coiling in 2D~\cite{picardo}, studying the impact of bending stiffness in this case will help us better understand the results in 3D. Moreover, this also provides us with the opportunity to check whether extensibility is crucial for preferential sampling in 2D as suggested in \cite{picardo}, {\it i.e.}, whether it is essential for a chain inside a vortex to collapse to a tracer, or if it is sufficient for a long chain to simply coil into the vortex.

Figure~\ref{fig:2D} presents snapshots of the inextensible chains overlaid on the vorticity field, for a highly flexible (panel {\it a}), a moderately flexible (panel {\it b}), and an inflexible (panel {\it c}) case 
(see also the movies in \cite{Movflex2D}). 
We see that the highly flexible chain is coiled up by the vortices and stays almost entirely
within them. The inflexible rod-like chain, however, is unable to coil, and thus while some portions of the chain linger within vortices, the entire chain can never be entrapped by a vortex and preferential sampling weakens considerably. This effect of bending stiffness, which is exactly as anticipated in \cite{picardo}, is shown quantitatively by the PDFs of $\Lambda_c$ in figure~\ref{fig:2D}{\it d} (see \eqref{eq:OW} for the
definition of $\Lambda_c$).
An analogous behaviour is observed for inextensible, inertialess fibers, described by the local slender-body theory, in a two-dimensional turbulent flow~\cite{JB-Pvt}

%

Figure~\ref{fig:2D}{\it d} also allows us to address the role of extensibility in 2D, by comparing the PDF of $\Lambda_c$ for an extensible, freely-jointed, chain ($\WiE = 20$, studied in \cite{picardo}) with the PDF for the inextensible, highly flexible chain ($\WiB = 0.4$). The extensible chain has a small equilibrium length of $0.04$ and a maximum length $L_m = 1.25$, which equals the fixed length $L$ of the inextensible chain. We see that while there is a small reduction of the degree of preferential sampling when the chain is unable to collapse to a tracer-like object, the long inextensible chains are still able to occupy vortices effectively (Figure~\ref{fig:2D}{\it a}), provided of course that they are flexible. Therefore, extensibility is not a crucial ingredient for preferential sampling, either in 2D or in 3D (shown in section~\ref{sec:inextensible}). However, it does have a more significant effect in 2D because extensible chains can collapse to tracers inside 2D vortices, whereas this is not possible inside axially-stretched 3D vortex tubes.

\begin{figure}[t]
\centering\includegraphics[width=\textwidth]{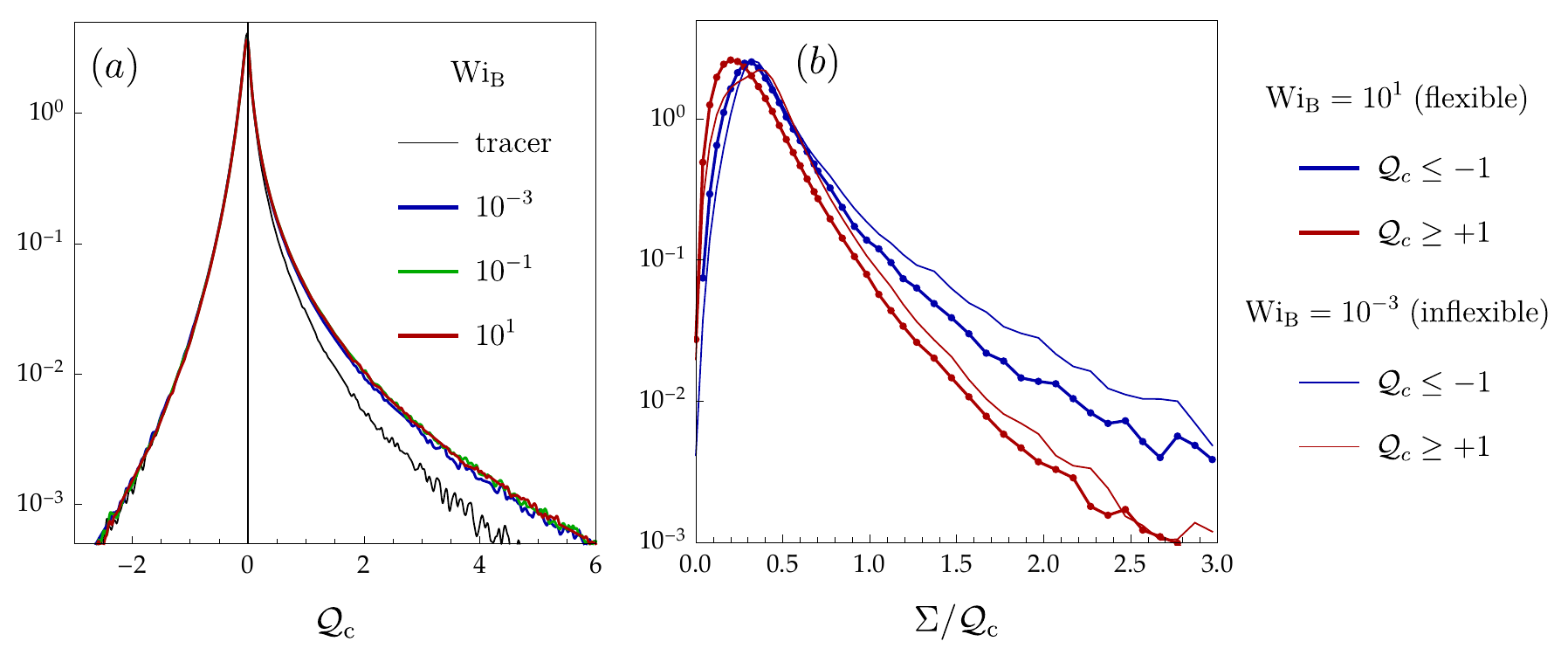}
\caption{(\textit{a}) PDF of $\Qc$ for a tracer
and for an inextensible chain ($L = 20 \eta$) with bending stiffness varying from that of a highly flexible chain ($\WiB = 10^1$) to that of a rod-like inflexible one ($\WiB = 10^{-3}$).
(\textit{b}) PDF of the standard deviation $\Sigma$ of the values of ${\cal Q}_j$ sampled by the beads of a chain, conditioned on the center-of-mass being in a vortical ($\Qc \ge +1 $) or straining region ($\Qc \le -1 $). Results are shown for the cases of a highly flexible ($\WiB = 10^1$) and inflexible ($\WiB = 10^{-3}$) chain of panel {\it a}.}
\label{fig:bending}
\end{figure}

Returning to the effect of the bending stiffness, we have shown above 
that in 2D it strongly affects the preferential sampling of vortical
regions because chains must coil in order to enter vortices. 
We now examine the effect of the bending stiffness in three
dimensions. Figure~\ref{fig:bending}\textit{a} shows the PDF of $\Qc$ for a long inextensible chain ($L = 20 \eta$) with different values of $\WiB$, corresponding to a highly flexible ($\WiB = 10^1$), a moderately flexible ($\WiB = 10^{-1}$), and a rod-like inflexible ($\WiB = 10^{-3}$) chain. The PDF clearly does not vary with $\WiB$,
which is an unequivocal indication that, in three dimensions,
the bending stiffness has no effect on preferential sampling. 
This fact provides further confirmation of our explanation of
preferential sampling of vortical regions in 3D turbulence. If the main
mechanism is the alignment of the chain with the axis of the tubular
structures on which vorticity concentrates, the chain does not
need to coil in order to reach and keep that configuration. It is
then natural that the bending stiffness does not affect 
preferential sampling in three dimensions.

All our results, thus far, point to the ability of chains to thread vortex tubes and follow them by aligning with the vortex axis. In contrast, no such mechanism exists for chains to reside inside straining zones which have a complex
geometry and evolve non-locally, as pointed out earlier in section~\ref{sect:extensible}. Indeed this difference between vortical and straining regions forms the basis for our explanation of preferential sampling. Therefore, it is important to support this claim with some direct evidence.
To this end, we denote by $\mathcal{Q}_j$ the value of $\mathcal{Q}$
at the position of the $j$-th bead, $\bm x_j$.
We then consider the mean $\mu$ and standard deviation $\Sigma$ of $\mathcal{Q}_j$ over the
$N_b$ beads of the chain: 
\begin{equation}
\mu = \left\langle \mathcal{Q}_j\right\rangle_b \quad {\rm and} \quad \Sigma = \big\langle \left(\mathcal{Q}_j-\mu\right)^2\big\rangle_b^{1/2}
\end{equation}
with $\langle\cdot\rangle_b$ denoting the average over $j=1,\dots,N_b$.
Figure~\ref{fig:bending}\textit{b} compares the conditional PDFs of $\Sigma$ for chains in rotational (${\cal Q}_c \geq +1$) and straining (${\cal Q}_c \leq -1$) regions, for both flexible and inflexible (inextensible) chains. In both cases, we see that $\Sigma$ is typically greater
when the centre of mass of the chain is in a high-strain region. This finding confirms that
it is generally easier for a chain, even if it is rod-like, to reside within vortex tubes
than within straining regions.





\section{Concluding remarks}
We have shown that a long bead-spring chain---a simple model for filamentary objects---preferentially samples vortical regions of both 2D and 3D turbulent flows. This behaviour is exhibited by a long inextensible chain, as well as an extensible one that has a small equilibrium length but which is stretched out to inertial-range scales by the flow. The key difference between the underlying mechanisms in 2D and 3D is revealed by the contrasting behaviour of inflexible chains. In 2D, where the chain must coil in order to be trapped within vortices, bending stiffness inhibits preferential sampling. In 3D, however, an inflexible chain exhibits the same sampling behaviour as a fully-flexible one, because the chain follows tubular vortices by aligning with their axes, and neither bending nor extensibility is essential for this to occur.

The extent of preferential sampling is much stronger in 2D that in 3D turbulence (compare the difference between the PDFs for a tracer and the chain in figure~\ref{fig:2D}{\it d} and figure~\ref{fig:bending}{\it a}). This is because 2D coherent vortices have much longer lifetimes than 3D vortex tubes, allowing the chain to spend a much larger fraction of time inside the former. On the other hand, the preferential sampling phenomenon is more robust in 3D, as it persists regardless of the extensibility or flexibility of the chain.

Filamentary objects in turbulent flows are encountered in several physical situations, from fibres in the paper industry to micro-plastics and algae in oceanic environments. The understanding of the sampling behaviour of individual chains developed here sets the stage for future studies of the transport, dispersion, and settling of filaments in turbulent suspensions, for which, however, additional physical effects such as intra-chain and inter-chain hydrodynamic interactions must be considered. 

\begin{acknowledgments} 
We thank J. Bec for several useful discussions on this topic. DV acknowledges his Associateship with the International Centre for Theoretical Sciences, Tata Institute of Fundamental Research, Bangalore 560089, India. JRP, SSR, and DV acknowledge the support of the Indo-French Centre for Applied Mathematics (IFCAM). 
JRP acknowledges funding from the IITB-IRCC seed grant. SSR and RS acknowledge support of the DAE, Govt. of India, under project no. 12-R\&D-TFR-5.10-1100. SSR also acknowledges DST (India) project MTR/2019/001553 for support.
\end{acknowledgments} 



\end{document}